\documentclass[twocolumn,showpacs,preprintnumbers,amsmath,amssymb]{revtex4}
\usepackage[T1]{fontenc}
\usepackage[latin1]{inputenc}
\usepackage{amsmath,amssymb,graphicx}
\makeatletter
\def\det{\mathop{\operator@font det}\nolimits}
\def\Re{\mathop{\operator@font Re}\nolimits}
\def\Im{\mathop{\operator@font Im}\nolimits}
\makeatother

\begin{document}

\title{Are better conducting molecules more rigid?}

\author{Young-Ho Eom}
\author{Hawoong Jeong}
\affiliation{Department of Physics, Korea Advanced Institute of
Science and Technology, Daejeon, 305-701, Korea}

\author{Juyeon Yi}
\email[e-mail: ]{jyi@pusan.ac.kr} \affiliation{Department of
Physics, Pusan National University, Busan
            609-735, Korea}

\author{Henri Orland}
\affiliation{Service de Physique Th\'{e}orique, CEA-Saclay, 91191
Gif-sur-Yvette Cedex, France}

\date{March. 22, 2006}

\begin{abstract}
We investigate the electronic origin of the bending stiffness of
conducting molecules. It is found that the bending stiffness
associated with electronic motion, which we refer to as
electro-stiffness, $\kappa_{e}$, is governed by the molecular
orbital overlap $t$ and the gap width $u$ between HOMO and LUMO levels
, and behaves as $\kappa_{e}\sim t^{2}/\sqrt{u^2+t^{2}}$. To
study the effect of doping, we analyze the electron filling-fraction 
dependence on $\kappa_{e}$ and show that doped molecules
are more flexible. In addition, to estimate the contribution of $\kappa_{e}$
to the total stiffness, we consider molecules
under a voltage bias, and study the length contraction ratio as
a function of the voltage. The molecules are shown to be
contracted or dilated, with $\kappa_{e}$
increasing nonlinearly with the applied bias.
\end{abstract}

\pacs{}

\maketitle

Conducting polymers have been extensively studied over the last 
decades because of their vast electronic device applications.
They have $\pi$-orbital overlapping along a conjugated backbone and a gap
between the highest filled and the lowest unfilled bands, forming
a band structure similar to inorganic semiconductors. When an
electron is removed or added, the conductivity is greatly enhanced, allowing these
semiconductors to be utilized as organic electronic device. 
On the other hand, conducting polymers are much more flexible 
than semiconducting
solids, while their electronic functions are very similar. 
%%%%%%%%%%%%%%%%%%%%%%%%%%%%%%%%%%%%%%%%%%%%%%%%%%%%%%%%%%%%%%%%%%%%%%
This property has led to studies for exploring
their mechanical deformations such as bending and
expansion/contraction driven by electric triggering. It is known that
the oxidation causes the expansion or the contraction of conducting
polymers, depending on whether ions are inserted or
eliminated~\cite{kaneto1}. Also, chemical doping drives
electromechanical deformation by delocalizing the
$\pi$-electrons~\cite{kaneto2,watanabe}.

The structural stiffness of a material is determined by many
factors, such as atomic binding energies, the nature of
molecular bonding, inter-polymer adhesion, to mention a few. 
It is worthwhile to
note that the orbital stack would be an additional source for
mechanical rigidity, because in order to achieve better electronic
conduction, structural deformations leading to a loss in orbital overlap
are less favored. 
Despite this great diversity of causes for the 
mechanical stiffness of materials, in this work, we 
present the first step to
relate electronic properties of these materials with their mechanical stiffness. 
This is
essential not only to relate the electronic origin of stiffness with
chemistry-based factors but also to grasp the mechanical response
to electric signal which differs from polymer to polymer. 
For
instance, the extension ratio of Polyalkylthiophene~(PAT) is
larger than that of Polypyrrole~(PPy)~\cite{kaneto}. It is interesting to 
relate it to the difference in their electronic band
structures. Although they have comparable orbital overlap, the gap
between HOMO and LUMO levels in PAT is about half of that in
PPy~\cite{hutchison}. This indicates that not only orbital overlap
but also band gap comes into play for mechanical stiffness.
%%%%%%%%%%%%%%%%%%%%%%%%%%%%%%%%%%%%%%%%%%%%%%%%%%%%%%%%%%%%%%%%%%%%%%%%%%%
\begin{figure}[b]
\centerline{\includegraphics[width=.95\columnwidth]{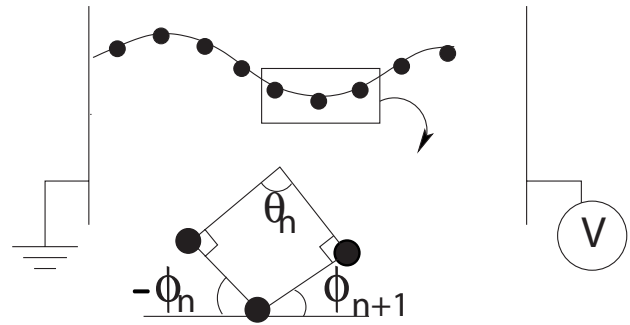}}
\caption{Schematic picture of a conducting polymer between
two electrodes. The one with the anchored polymer is grounded and the
other is fixed at voltage $V$. The polymer
configurations are discretized in a way such that the bond angle and the mean position
of two adjacent beads define the tangential angle $\phi_{n}$ and
position of $n$th monomer, respectively.}
\end{figure}
%%%%%%%%%%%%%%%%%%%%%%%%%%%%%%%%%%%%%%%%%%%%%%%%%%%%%%%%%%%%%%%%%%%%%%%%%%%

In this work, we investigate the effect of the electronic properties of 
conducting polymers on their conformational stiffness. 
This enables us to
understand mechanical responses as well as to estimate the
electronic contribution to the total structural rigidity. 
For this purpose, 
we model conducting polymers as a one-dimensional chain
composed of $N$-monomer where inter-monomer hopping of electrons
is allowed via $\pi$-orbital stacking. 
More specifically, to investigate
the bending rigidity, we assume the hopping strength to depend
on the angular configuration of monomers \cite{Hone}. Furthermore, to simulate the
HOMO-LUMO gap, an alternating on-site potential is included in the
Hamiltonian. Following a semi-classical approach, we trace over
the electronic degree of freedom and obtain the effective
potential for the angle deformations. We find that the bending
stiffness associated with electronic properties, which we refer to as
electro-stiffness, $\kappa_{e}$, is governed by the molecular
orbital, $t$ and gap width between HOMO and LUMO level, $u$, and scales 
as
$\kappa_{e}\sim t^{2}/\sqrt{u^2+t^{2}}$. Further by analyzing
the electron-filling fraction dependence on $\kappa_{e}$, we show that
doping would make the molecules more flexible. To evaluate
the specific contribution of $\kappa_{e}$ to the total rigidity, we
consider molecules under a voltage bias and examine the length
contraction as a function of the bias. It turns out that the
applied voltage alters the electro-stiffness, resulting in the
length contraction~(dilation) of the molecules.

We consider a conducting molecule in the presence of an external
electric field. The Hamiltonian for the electrons responsible for the
conducting behavior is taken as
\begin{equation}
{\cal H}_{e}=-\sum_{n}t_{n,n+1}(c_{n}^{\dagger}c_{n+1}+\mbox{h.c})
+\sum_{n}[(-1)^{n}u-vx_{n}]c^{\dagger}_{n}c_{n}. \label{elham}
\end{equation}
Here the inter-site hopping integral $t_{n,n+1}$ is determined by
the degree of $\pi$-orbital overlap, and is maximum when the molecule is
in a straight form. 
When the molecule is not in the straight conformation, 
the overlap is decreased, which tends
to suppress electron
hopping. We incorporate this fact by introducing an
angle-dependence in the hopping integral as
$t_{n,n+1}=t\cos(\phi_{n+1}-\phi_{n})\equiv t\cos\theta_{n}$. For
$\theta_{n}=0$, the hopping parameter is spatially uniform and
maximized by parallel orbital arrangement. In the second term we
introduce an alternating on-site potential, yielding a gap
whose width is determined by $u$. This enables us to investigate
semiconducting molecule having a gap between HOMO and LUMO level.
Also, the on-site potential is position dependent due to the
applied voltage $V$ and $v=|e|Va/L$ is the voltage drop across one
monomer with electrode spacing $L$(see Fig.~1). Taking the monomer
spacing as $a$, we write
$x_{n}=\sum_{i=0}^{n}\cos\phi_{i}-(1/2)\cos\phi_{n}$.

The Hamiltonian contains several electronic factors that contribute to
the structural rigidity: (i)the electron hopping favored by a
straight configuration; (ii)the band gap making the molecules less
conducting and tending to diminish the effect of the factor (i);
(iii) the applied voltage bias inducing length contraction or
extension. 
The rigidity from electronic origin,
$\kappa_{e,n}$, can be obtained by
\begin{equation}
\kappa_{e,n}=-\frac{k_{B}T}{2}\frac{\partial^{2}}{\partial
\theta_{n}^{2}}\ln \mbox{Tr}_{\{c,c^{\dagger}\}}\exp(-\beta {\cal
H}_{e}),\label{exact}
\end{equation}
which can be site-dependent when boundary effects are
considered. 
In fact, the rigidity also has  contributions from molecular bonding
and atomic binding potentials, which we denote as $\kappa_{m}$, and
the potential governing the structural rigidity is written as
\[
{\cal H}_{m}=\sum_{n}\kappa_{n}\theta_{n}^{2},
\]
where $\kappa_{n}=\kappa_{e,n}+\kappa_{m}$.
%%%%%%%%%%%%%%%%%%%%%%
Before proceeding to the numerical evaluation of Eq.~(\ref{exact}), we study the small bending of the
molecules in the absence of a voltage drop. Expanding the angle
to quadratic order, we define
\begin{eqnarray}
{\cal H}_{e}^{(0)}&=&-t\sum_{n}(c_{n}^{\dagger}c_{n+1}+\mbox{h.c})
+u\sum_{n}( -1)^{n}c_{n}^{\dagger}c_{n} \\ \nonumber {\cal
H}_{e}^{(1)}&=&\frac{t}{2}\sum_{n}\theta_{n}^{2}
(c_{n}^{\dagger}c_{n+1}+\mbox{h.c}),
\end{eqnarray}
%\begin{eqnarray}
%{\cal H}_{e}^{(0)}&=&-t\sum_{n}(c_{n}^{\dagger}c_{n+1}+\mbox{h.c})
%+u\sum_{n}( -1)^{n}c_{n}^{\dagger}c_{n} \\ \nonumber {\cal
%H}_{e}^{(1)}&=&\frac{t}{2}\sum_{n}\theta_{n}^{2}
%(c_{n}^{\dagger}c_{n+1}+\mbox{h.c})+e\mathrm{E}\sum_{n}x_{n}c^{\dagger}_{n}c_{n},
%\end{eqnarray}
where the position of the $n$th monomer for small $\phi$'s is
denoted by $x_{n}$. Assuming ${\cal H}_{e}^{(1)}$ to be a small
perturbation, we
can write $-k_{B}T\ln \langle e^{-\beta{\cal H}_{e}^{(1)}}\rangle
_{0} \approx \langle {\cal H}_{e}^{(1)}\rangle_{0}$,
 where $\langle X
\rangle_{0}={\cal Z}_{0}^{-1}\mbox{Tr} X e^{-\beta {\cal
H}_{e}^{(0)}}$ with ${\cal Z}_{0}=\mbox{Tr}e^{-\beta {\cal
H}_{e}^{(0)}}$. It is convenient to work in the Fourier space:
$c_{n}=N^{-1/2}\sum_{n}e^{ikn}c_{k}$. The Hamiltonian ${\cal
H}_{e}^{(0)}$ can then be straightforwardly diagonalized by
the canonical transformation
$c_{k}=\cos\chi_{k}a_{k,+}+\sin\chi_{k}a_{k,-}$ and
$c_{k+\pi}=-\sin\chi_{k}a_{k,+}+\cos\chi_{k}a_{k,-}$ as
\[
{\cal
H}_{e}^{(0)}=\sum_{k,\alpha=\pm}\lambda(k)a_{k,\alpha}^{\dagger}a_{k,\alpha},
\]
where $\lambda(k,\alpha)=\alpha\sqrt{\epsilon^{2}(k)+u^{2}}$ with
$\epsilon (k)=-2t\cos k$ and $\tan 2\chi_{k}=-u/\epsilon(k)$
and $\alpha = \pm 1$.
Similarly we can write ${\cal H}_{e}^{(1)}$ in terms of the
diagonalizing basis, with off-diagonal components. In
evaluating $\langle {\cal H}_{e}^{(1)}\rangle_{0}$, however,
since ${\cal H}_{el}^{(0)}$ is quadratic in the $c$'s, the only
non-vanishing contributions can be easily traced and we get
\begin{equation}
\kappa_{e}=\frac{1}{N}\sum_{k,\alpha =\pm 1} \frac{-\epsilon
^{2}(k)}{\alpha\sqrt{u^{2}+\epsilon^{2}(k)}}\langle {\cal
N}_{k,\alpha}\rangle _{0} \label{electrostiff}
\end{equation}
with the mean number of particles
occupying the $\alpha=+1$ and $-1$ bands being determined by
\begin{equation}
\langle {\cal N}_{k,\alpha}\rangle _{0}=\langle
a^{\dagger}_{k,\alpha}a_{k,\alpha}\rangle_{0}=[e^{\beta(
\lambda(k,\alpha)-\mu)}+1]^{-1}.
\end{equation}
%%%%%%%%%%%%%%%%%%%%%%%%%%%%%%%%%%%%%%%%%%%%%%%%%%%%%%%%%%%%%%%%%%%%%%%%%%%
\begin{figure}[t]
\centerline{\includegraphics[angle=-90,width=.95\columnwidth]{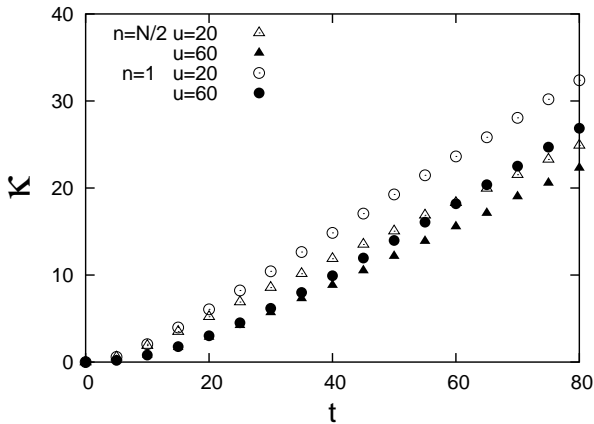}}
\caption{The electro-stiffness $\kappa_{e}$ vs the hopping
integral for $N=40$ in the absence of a voltage bias. The values
of $\kappa_{e}$ depend on the monomer position, those near the
boundaries, e.g. $n=1$ being smaller than those at $n=N/2$. For a given $n$,
$\kappa_{e}$ is estimated for $u=20$ and $u=60$ to show that
an increase in the gap width suppresses the stiffness. The energy
scales such as $\kappa_{e}, t$ and $u$ are measured
in units of thermal energy at room temperature,
$k_{B}T=0.025$~eV, throughout this paper.} 
\end{figure}
%%%%%%%%%%%%%%%%%%%%%%%%%%%%%%%%%%%%%%%%%%%%%%%%%%%%%%%%%%%%%%%%%%%%%%%%%%%
Let us first consider when the system is half-filled so that
$\langle {\cal N}_{k,+}\rangle =0$. From Eq.~(\ref{electrostiff}),
it is clear that as far as the electronic contribution to the
bending stiffness is concerned, the hopping integral plays a
dominant role: while the denominator in Eq.~(\ref{electrostiff})
characterizes the band width, the numerator is proportional to
$t^{2}$. On the other hand, for molecules having comparable
hopping strengths, those with large band gap would be more
flexible. The numerical evaluation of Eq.~(\ref{exact}) has been
performed and the resulting stiffness is presented in Fig.~2. It
is in good agreement with the perturbative
result, Eq.~(\ref{electrostiff}). While the
analytic expression for the stiffness can be easily
obtained for an infinitely long polymer,
a finite-sized polymer has boundary effects that 
are characterized by a site-dependent stiffness. 
As shown in Fig.~2, the stiffness is weaker in the 
polymer bulk than in the ends, indicating that
when a force is applied, the bending of the polymer would be more localized  in the bulk rather than being uniform all over.

%As shown in Fig.~2,
%at polymer end the stiffness is weaker than at bulk, indicating
%that when a force is applied, rather than taking uniform curvature
%polymers would localize the bend near ends.

%%%%%%%%%%%%%%%%%%%%%%%%%%%%%%%%%%%%%%%%%%%%%%%%%%%%%%%%%%%%%%%%%%%%%%%%%%%
\begin{figure}[t]
\centerline{\includegraphics[angle=-90,
width=.95\columnwidth]{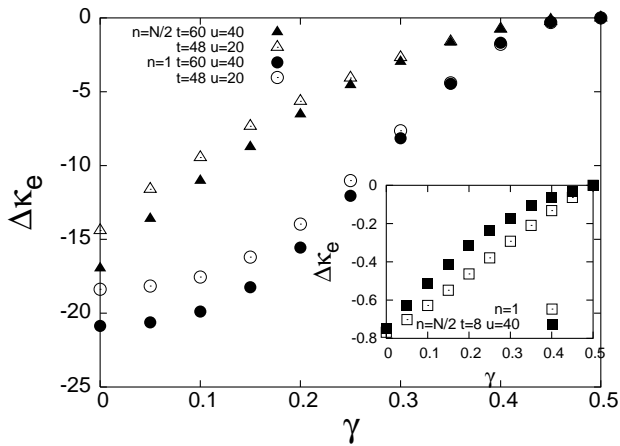}} \caption{The stiffness
difference $\Delta\kappa_{e}$ vs the electron filling fraction
$\gamma$ for N = 40 in the absence of a voltage bias: for a
given $n$, $\kappa_{e}$ is estimated for $t=60$, $u=40$ (PPy) and
$t=48$, $u=20$ (PAT) to show that the increase of the gap width suppresses
the stiffness. The inset shows $\kappa_{e}$ for a DNA molecule with
$t=8$, $u=40$. Here $\kappa_{e}, t$ and $u$ are in units of thermal
energy at room temperature, $k_{B}T=0.025$~eV}.
\end{figure}
%%%%%%%%%%%%%%%%%%%%%%%%%%%%%%%%%%%%%%%%%%%%%%%%%%%%%%%%%%%%%%%%%%%%%%%%%%%%%
Even more interesting is the filling factor dependence of
$\kappa_{e}$. When hole/particles are introduced in the system, the
electrostiffness becomes weaker than that for a half-filled system,
as displayed in Fig.~3. We evaluated the stiffness difference
$\Delta \kappa_{e} \equiv \kappa_{e}(\gamma)-\kappa_{e}(1/2)$,
where $\gamma$ denotes the filling fraction as $\gamma = N_{e}/N$.
This is in good agreement with the experimental observation that
the bending rigidity of Polyurethane film is enhanced by salt
doping~\cite{watanabe}. This becomes obvious from
Eq.~(\ref{electrostiff}). When the system is half-filled,
$\alpha=(+1)$ band would be empty so that in
Eq.~(\ref{electrostiff}), the contribution to the stiffness is
solely due to the $\alpha=(-1)$ band. On the other hand, when holes are
doped, the $\alpha=(-1)$ band becomes partially filled, and
the corresponding reduction in $\langle {\cal N}_{k,-}\rangle$ results in a decrease of $\kappa_{e}$.
%%%%%%%%%%%%%%%%%%%%%%%%%%%%%%%%%%%%%%%%%%%%%%%%%%%%%%%%%%%%%%%%%%%%%%%
\begin{figure}[t]
\centerline{\includegraphics[angle=-90,width=.95\columnwidth]{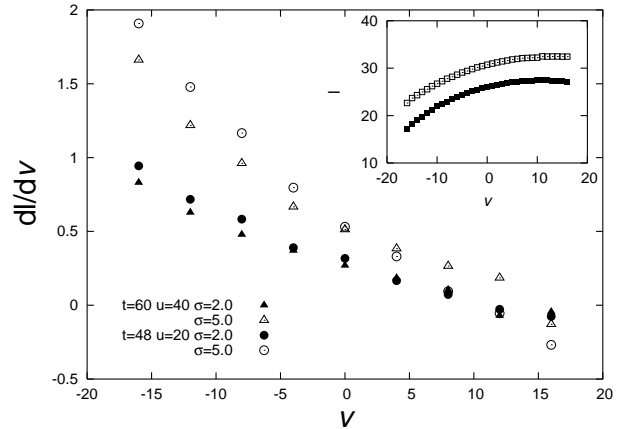}}
\caption{The derivative of the length contraction ratio $\ell$  vs the
voltage bias $v$ for various parameter sets. Here $\sigma$
represents the ratio $\sigma= \frac{\kappa_m}{\kappa_e}$. The inset shows the length contraction
ratio $\ell$ vs $v$ for $\sigma =2$. The square symbols and the filled
squared symbols represent $t=60, u=40$ and $t=48, u=20$,
respectively. Here $\kappa_{e}, t$ and $u$ are in units of thermal
energy at room temperature, $k_{B}T=0.025$~eV}.
\end{figure}
%%%%%%%%%%%%%%%%%%%%%%%%%%%%%%%%%%%%%%%%%%%%%%%%%%%%%%%%%%%%%%%%%%%%%%%%
The electrostiffness depends also on the applied electric field
which contributes to the energy of the system via a coupling to the charges of the electrons: when $v \ll t$ and $u$, we may 
approximate the energy due to the field as $E_{f}=-v\sum_{n} x_{n}\langle
c_{n}^{\dagger}c_{n}\rangle $. When a positive voltage bias is
applied, the molecule increases its length, and hence the effective
stiffness increases. For a negative voltage bias, the molecule tends
to contract, resulting in the reduction of its stiffness.
Investigating the length deformation as a function of the electric field
is thus very useful to evaluate the contribution of $\kappa_{e}$
to the total rigidity $\kappa$. As we mentioned earlier, the
structural rigidity comes not only from those electronic degrees
of freedom but also from the molecular binding potentials. 
Since for the
latter, the deformation is presumably rather insensitive to the applied electric
potential, the length deformation caused by the electric field would
directly relate the contribution of $\kappa_{e}$ to the total
rigidity. To see this more clearly, let us define the length
contraction (extension) ratio $\ell$ as
\begin{equation}
\frac{1}{\ell}\equiv 1-\frac{L}{L_{c}}\approx \frac{1}{2}\langle
\phi^{2}\rangle
\end{equation}
with $L_{c}$ being the contour length of the polymer, and $\langle
\phi^{2}\rangle$ being the average angle fluctuation per monomer,
given by $\langle \phi^{2}\rangle = \mbox{Tr}_{\{\phi\}}e^{-\beta
\mathcal{H}_{m}}\sum_{i}\phi_{i}^{2}/
\mbox{Tr}_{\{\phi\}}e^{-\beta\mathcal{H}_{m}}$. Here, disregarding
the site-dependence in $\kappa_{e,n}$ (which as we saw is uniform
, except for a few boundary
sites),
and writing $\mathcal{H}_{m}=\sum_{\langle i,j\rangle
}\kappa(\phi_{i}-\phi_{j})^{2}$ where
$\kappa=\kappa_{m}+\kappa_{e}$ and $\kappa_{m}$ is the bending
rigidity due to molecular bonding. It is clear that
$\langle \phi^{2}\rangle \propto 1/\kappa$ and hence, $\partial
\ell/\partial v$, which is measurable in experiments, would
be simply related to the electric-field dependence of $\kappa_{e}$. 
Evaluating
$\ell$ as a function of the voltage bias for molecules having
different $\kappa_{m}$, e.g, $\sigma\equiv
\kappa_{m}/\kappa_{e} (v=0, 2, 5)$, we confirm in Fig.~4
that
$\partial\ell/\partial v$ is indeed not very sensitive to
$\kappa_{m}$.  
It is also shown that the
length contraction ratio increases nonlinearly with $v$ and its
derivative with respect to $v$ is positive. This clearly
demonstrates the expected feature that the molecules adjust their
length to the voltage, allowing for their use as
electro-mechanical switches.

In summary, the electronic origin of bending stiffness was
investigated. It was shown that the electro-stiffness,
$\kappa_{e}$, is governed by the molecular orbital overlap and the
gap width between HOMO and LUMO levels: molecules with wider band
width are more flexible. The electro-stiffness can be controlled by
molecular doping or by applying a voltage bias. Analyzing
the electron filling-fraction dependence on $\kappa_{e}$, we
showed that doping makes molecules more flexible. In addition,
we considered molecules under a voltage bias to extract the
$\kappa_{e}$ contribution to the total stiffness. 
%%%%%%%%%%%%%%%%%%%%%%%%%%%%%%%%%%%%%%%%%%%%%%%%%
In response
to the applied voltage, the molecules are contracted
or dilated with a very nonlinear increase of $\kappa_{e}$ 
with the applied bias.

To conclude this study, we mention the value of
$\kappa_{e}$ for a few molecules. For example, DNA
has an extremely narrow band width~($\approx 0.01\sim 0.04$
eV)~\cite{zhang} and its electrostiffness is estimated to be
roughly just a few $k_{B}T$. It is well-known that the persistence
length of ssDNA is about $\ell_{p}\sim 5$nm~\cite{dnaper}, which
is related to the bending energy by $\kappa =(\ell_{p}/a)k_{B}T$.
Taking the inter-base distance $a=3.4\AA$, $\kappa\approx 14
k_{B}T$, showing that  the contribution of $\kappa_{e}$ to
the stiffness is significant. For a dsDNA
$\ell_{p}\approx 50 nm$ and hence $\kappa$ is ten times bigger
than that for ssDNA~\cite{dnaper}, while the doubling of $\kappa_{e}$ cannot
account for the difference. This suggests that among the energetic factors
which govern dsDNA bending, the electronic motions via orbital
overlap is not as crucial as the electrostatic repulsion between
phosphate groups and the helical structured base
staking~\cite{nar}. On the other hand, the persistence length of
carbon nanotubes~(CNT) lies in a macroscopic range 
$\ell_{p}=0.1\sim 1 \mu$m~\cite{cntper}, which shows that the bending
rigidity of CNT is hundreds times larger than that of dsDNA.
Noting that the orbital overlap is $t\sim 2.5$ eV, and the band gap is small,
$u=0\sim 0.5 eV$~\cite{ijima}, we find
$\kappa_{e}\approx 10^{2}k_{B}T$, which shows the important contribution of 
electro-stiffness to the total stiffness of CNT. 
In addition, the HOMO and LUMO level of PPy and PAT can be simulated
by taking $u=1$~eV and $t=1.5$~eV, and $u=0.5$~eV and $t=1.2$~eV,
respectively~\cite{hutchison}. This leads to $\kappa_{e}\approx 34
k_{B}T$ for PPy, and $\kappa_{e}\approx 28 k_{B}T$ for PAT. Although no
direct measurement of the bending rigidity of these materials has been made,
this goes in the direction of showing that the latter is more
responsive than the former~\cite{kaneto}.

%H. Orland and D. Porath are greatly acknowledged for stimulating discussion
%and helpful comments on this manuscript.

\end{document}